\documentclass[prc,preprint,showpacs,preprintnumbers,amsmath,amssymb]{revtex4}

\usepackage[mathscr]{eucal}
\usepackage{graphicx}
\def\slr#1{\setbox0=\hbox{$#1$}           
   \dimen0=\wd0                                 
   \setbox1=\hbox{/} \dimen1=\wd1               
   \ifdim\dimen0>\dimen1                        
      \rlap{\hbox to \dimen0{\hfil/\hfil}}      
      #1                                        
   \else                                        
      \rlap{\hbox to \dimen1{\hfil$#1$\hfil}}   
      /                                         
   \fi}

\def\ksq{k^2}

\def\mytint#1{\!\int\!\!\frac{d^3\!{#1}}{(2\pi)^3}\,}
\def\gev#1{ GeV${}^{#1}$}
\def\be{\begin{eqnarray}}
\def\ee{\end{eqnarray}}

\renewcommand{\theequation}%
    {\arabic{section}.\arabic{equation}}
\makeatletter \@addtoreset{equation}{section} \makeatother

\begin{document}

\preprint{BCCNT: 02/121/318}

\title{Calculation of Temperature-Dependent Hadronic Correlation Functions
of Pseudoscalar and Vector Currents}

\author{Bing He}
\author{Hu Li}
\author{C. M. Shakin}
 \email[email:]{casbc@cunyvm.cuny.edu}
\author{Qing Sun}

\affiliation{
 Brooklyn College of the City University of New York\\ Brooklyn, New York
11210 }%

\date{December, 2002}

\begin{abstract}
We make use of the Nambu-Jona-Lasinio (NJL) formalism and
real-time finite-temperature field theory to calculate hadronic
current correlation functions in the deconfined phase of quantum
chromodynamics (QCD). Here, we consider both pseudoscalar and
vector currents. Since the method used in the lattice analysis to
calculate the spectral functions requires assumptions about the
likelihood of a particular form for the spectral function, we
believe our calculations will be useful to researchers who wish to
calculate hadronic current correlation functions at finite
temperature using lattice-based methods. Our model makes use of
temperature-dependent coupling constants for the NJL model. We
present various arguments that such temperature dependence is
necessary, if the results of the model are to be consistent with
what is known concerning QCD thermodynamics.
\end{abstract}


\maketitle

\section{INTRODUCTION}

In recent years we have seen a great deal of interest in the
properties of dense matter, with particular attention given to
diquark condensation and color superconductivity [1]. Since it is
difficult to study the properties of dense matter in lattice
simulations of QCD [2], the Nambu-Jona-Lasinio (NJL) model and
closely related instanton-based models have been used in such
studies. We have become interested in a possible density
dependence of the coupling constants of chiral Lagrangian models,
since density dependence of the coupling parameters could affect
the conclusions drawn from the studies of dense matter. We have
introduced density-dependent coupling constants for the NJL model
in earlier works [3, 4], and have presented some arguments that
such density dependence may be necessary [4]. However, it is much
easier to discuss the temperature dependence of NJL coupling
parameters, rather that the density dependence, since a good deal
is known concerning finite-temperature QCD thermodynamics [5]. In
particular, the study of a gluon gas at high temperature suggests
that the system is well described as weakly interacting for $T
\gtrsim6 T_c$, where $T_c$ is the temperature of the
confinement-deconfinement phase transition. (See Fig.1.3 of Ref.
[5].)

In this work we suggest that rather straightforward calculations
of hadronic current correlation functions at finite temperature
can provide information concerning a possible temperature
dependence of the NJL coupling parameters. We perform calculations
of hadronic current correlation functions for pseudoscalar and
vector currents in the range $1.2 \leqslant T/T_c \leqslant5.88$.
We make use of two models for the temperature dependence of the
NJL coupling parameters. For model 1, we use
$G(T)=G[1-0.17(T/T_c)]$, which was the form used in our previous
studies of meson properties [6] and hadronic current correlation
functions [7] at finite temperature. In this work we also
introduce a model 2, for which $G(T)=G[1-0.0289(T/T_c)^2]$. In
both cases, $G(T)=0$ for $T/T_c=5.88$. For values of $T/T_c>5.88$
we put $G(T)=0$ for both models.

For the sake of completeness we present the Lagrangian of a
generalized NJL model that we have used in our studies of meson
properties at finite temperature and density \be {\cal L}=&&\bar
q(i\slr
\partial-m^0)q +\frac{G_S}{2}\sum_{i=0}^8[
(\bar q\lambda^iq)^2+(\bar qi\gamma_5 \lambda^iq)^2]\nonumber\\
&&-\frac{G_V}{2}\sum_{i=0}^8[
(\bar q\lambda^i\gamma_\mu q)^2+(\bar q\lambda^i\gamma_5 \gamma_\mu q)^2]\nonumber\\
&& +\frac{G_D}{2}\{\det[\bar q(1+\gamma_5)q]+\det[\bar
q(1-\gamma_5)q]\} \nonumber\\
&&+ {\cal L}_{conf}\,. \ee Here, $m^0$ is a current quark mass
matrix, $m^0=\mbox{diag} \,(m_u^0, \,m_d^0, \,m_s^0)$. The
$\lambda^i$ are the Gell-Mann (flavor) matrices. Here,
$\lambda^0=\sqrt{2/3}\mathbf{\,1}$ with $\mathbf{\,1}$ being the
unit matrix. The fourth term on the right-hand side of Eq. (1.1)
is the 't Hooft interaction. Finally, ${\cal L}_{conf}$ represents
the model of confinement we have used in our work.

In order to specify the parameters in Eq. (1.1) we may refer to
Ref [8]. There, a quite detailed fit was made to the properties of
the $\eta$ mesons. Our analysis yields fits the first four
experimentally known levels and provides excellent fits to the
mixing angles and decay constants of the $\eta$(547) and
$\eta\prime$(958). That work led to the specification of
$G_S=11.84$\gev{-2}. The 't Hooft interaction strength, $G_D$, was
taken to be in the range $-220$\gev{-5} $\leq G_D\leq -180$
\gev{-5} when calculating the properties of the $\eta$ mesons. We
also used $G_V=13.0$\gev{-2} in Ref.[8]. (It is worth noting that
Hatsuda and Kunihiro [9] have used a sharp moment cutoff of
$\Lambda_3=0.6314$ GeV and a value of $G_D=185.1$\gev{-5}. Their
value of $G_S$ is smaller than ours, since they do not include a
model of confinement in their analysis.)

In terms of these parameters we can introduce effective coupling
constants for states with particular quantum numbers. The
effective coupling parameter involves the vacuum condensates
$\alpha=<0\vert \bar uu\vert 0>$, $\beta=<0\vert \bar dd\vert 0>$
and $\gamma=<0\vert \bar ss\vert 0>$. For the
pseudoscalar-isovector states, we have [9] \be
G_P^{eff}=G_S+\gamma\frac{G_D}{2}\,. \ee We take
$\alpha=\beta=-(0.241 \mbox{GeV})^3$ and $\gamma=-(0.258
\mbox{GeV})^3$ [10]. If we put $G_D=-190$\gev{-5}, we find
$G_P=13.47$ \gev{-2}.

In Ref. [6] we presented relativistic random phase approximation
(RPA) calculations for the pion, kaon and several other mesons. It
was found that in the case of pion, that the use of
$G_P^{eff}=13.49 $\gev{-2} gave a pion energy of 140 MeV at $T=0$.
It is gratifying to see that an independent calculation gives rise
to essentially the same effective coupling constant as that
obtained from the parameters determined in Ref.[8].

If we consider the scalar-isovector states, the effective coupling
constant is \be G_S^{eff}=G_S-\gamma\frac{G_D}{2}\,, \ee which,
for $G_S=-11.84$\gev{-2}, $G_D=190.0$\gev{-5} and
$\gamma=-(0.258\mbox{GeV})^3$, yields $G_S=10.21$\gev{-2}. When we
compare $G_P^{eff}$ with $G_S^{eff}$, we see that we can expect
the resonant enhancement of the spectral functions for the
pseudoscalar-isovector channel will be larger in magnitude than
that of the spectral function of the scalar-isovector channel. On
the other hand, since the 't Hooft interaction does not affect the
vector or axial-vector coupling constants, these coupling
constants would be equal in our model, leading us to expect quite
similar spectral functions for these two channels.

In our study of the $\eta$ mesons we used $G_V=13.00$ \gev{-2}.
However,the results for the $\eta$ mesons are not particularly
sensitive to that parameter. In earlier work, with a sharp cutoff
for the NJL model of $\Lambda_3=0.622$ GeV, we used $G_V=12.46$
\gev{-2} to fit the energy of the $\omega$(782) [11]. In other
calculations we have used $G_V=11.46$\gev{-2}. In this work we
present the results for the latter value. (We have checked that
the spectral functions calculated with either $G_V=12.46$\gev{-2}
or $G_V=11.46$\gev{-2} differ by less than 4 percent.)

It was found in our various studies of meson properties, made
after we completed Ref. [11], that the use of a sharp cutoff did
not allow us to study radial excitations with a large number of
nodes. However, we found that a Gaussian regulator, $\exp[-\vec
k\,{}^2/\alpha^2]$, with $\alpha=0.605$ GeV, gave similar results
as that of the sharp cutoff $\Lambda_3=0.622$ GeV for the
low-lying states and also allowed us to describe radial
excitations with many nodes. Therefore, all our more recent
calculations have been made using the Gaussian regulator with
$\alpha=0.605$ GeV. (The coupling parameters obtained in Ref.[8],
for example, were determined with this Gaussian regulator, so that
any change of the regulator parameter, $\alpha$, would require
that we provide a new set of coupling parameters corresponding to
the new value of $\alpha$.)

We have recently reported results of our calculations of the
temperature dependence of the spectra of various meons [6]. These
calculations were made using our generalized NJL model which
includes a covariant model of confinement. We have presented
results for the $\pi$, $K$, $a_0$, $f_0$ and $K_0^*$ mesons in
Ref.[6]. In that work, temperature-dependent constituent quark
masses were calculated using the equation [12] \be
m(T)=m^0+2G_S(T)N_c\frac{m(T)}{\pi^2}\int_0^\Lambda
dp\frac{p^2}{E_p}\tanh(\frac{1}{2}\beta E_p)\,. \ee Here, $m^0$ is
the current quark mass, $G_S(T)$ is a temperature-dependent
coupling constant introduced in our model. (In Ref.[6] we used
model 1 for $G_S(T)$.) Here, $N_c=3$ is the number of colors,
$\beta=1/T$ and $E_p=\left [\vec p\;{}^2+m^2(T)\right]^{1/2}$.
Further, $\Lambda=0.631$ GeV is a cutoff such that $|\vec
p|\leq\Lambda$. We have put $\Lambda=0.631$ GeV, since that is the
cutoff that is often used when solving Eq. (1.4) [9]. We have also
used $m_u^0=5.5$ MeV and $m_s^0=120$ MeV. Since in our work
dealing with meson spectra, we have used $m_u=0.364$ GeV and
$m_s=0.565$ GeV as phenomenological parameters, we have put
$G_S(0)=2G$, with $G=5.691$\gev{-2}, in the notation of Ref.[12],
to reproduce those values. In Ref.[12] $G$ is one-half of the
$G_S$ defined in Eq.(1.1).

Since the use of temperature-dependent coupling constants is an
unusual feature of our work, we now describe some advantages that
follow from that choice. Our original introduction of temperature
dependence, $G(T)=G[1-0.17(T/T_c)]$, with $T_c=170$ MeV, had the
purpose of simulating the dynamical interactions which eliminate
pion or kaon condensation at relatively low temperatures. However,
we found when solving Eq.(1.4) for the temperature-dependent
constituent mass of the up quark, chiral symmetry was (partially)
restored at lower temperatures than when constant coupling
parameters were used. This may be seen in Fig.1, where we show the
results that follow from the use of a constant value of
$G(T)=5.691$\gev{-2}(in the notation of Ref.[12]) as a dotted
line. Here we have used $m^0=5.5$ MeV as the current quark mass.
For the solid and dashed curves in Fig.1 we have used
$G(T)=G[1-0.17(T/T_c)]$. For the sake of this discussion, let us
consider the reduction of the constituent mass to 50 MeV from 364
MeV as signal of the (partial) restoration of chiral symmetry. For
the dashed and solid curves, that restoration take place at about
$T_c=170$ MeV, while for the dotted curve the (partial)
restoration of chiral symmetry takes place at about 250 MeV. For
QCD, the transition temperature is about 150-170 MeV. Since we are
attempting to create a model that has some correspondence to QCD,
with dynamical quarks, we see that the temperature-dependent
coupling constants lead to the desired behavior.

The temperature dependence of the constituent mass in the case of
temperature-dependent coupling constants is such that we described
the mesonic confinement-deconfinement transition as taking place
at $T\approx T_c$ with $T_c=170$ MeV in Ref. [6]. In that work we
have studied the confinement-deconfinement transition for the
$\pi$, $K$, $a_0$, $f_0$ and $K_0^*$ mesons. We have checked that,
without the introduction of the temperature-dependent coupling
constants, the confinement-deconfinement transition would take
place at a significantly higher temperature.

A further advantage of the use of our temperature-dependent
coupling parameter is that the NJL interaction goes to zero for
$T=5.88T_c$.(It is put equal to zero for $T>5.88T_c$.) We suggest
that that is consistent with QCD thermodynamics, since it appears
that the system is weakly coupled for $T\geq 6T_c$ [5].

In Fig.2 we show the behavior of both the up and strange quark
constituent masses calculated with Eq.(1.4) using current masses
$m_u^0=5.50$ MeV and $m_s^0=120$ MeV. For the calculations made in
this work, we have used the temperature-dependent up (or down)
quark masses shown in Fig.2 for temperatures for which the quark
masses are quite small. Therefore, our results are insensitive to
variations of the mass values shown in Fig.2.

In calculating the constituent mass values we have neglected the
confining interaction. That interaction was taken into account in
our earlier Euclidean-space calculation of the quark self-energy
[13], which also included the effects related to the 't Hooft
interaction. We found that, to a good approximation, we could
neglect the confining and 't Hooft interactions, if we modified
the value of the NJL coupling constant, $G_S$, and we adopt that
approach when using Eq.(1.4).

 \begin{figure}
 \includegraphics[bb=0 0 280 235, angle=0, scale=1.2]{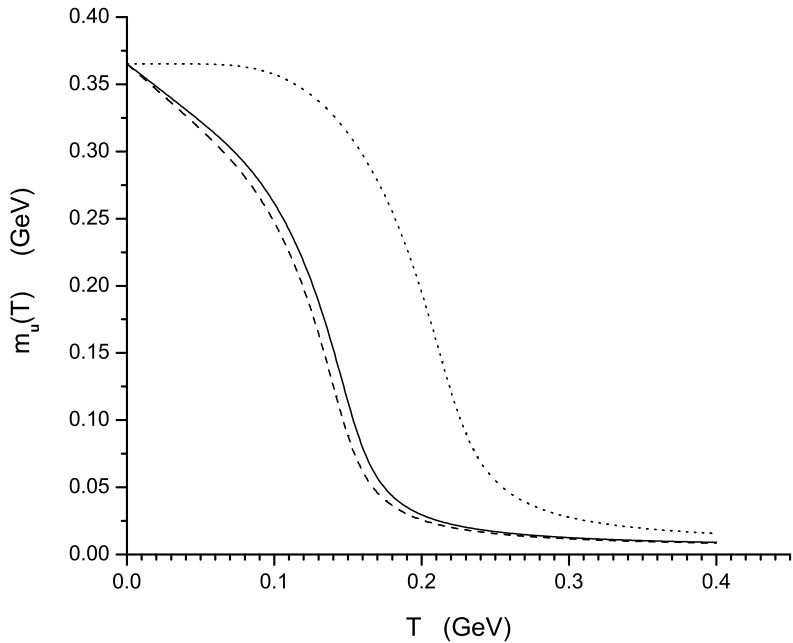}%
 \caption{We exhibit the solution for $m_u(T)$ obtained using Eq.(5.38) of Ref.[12],
 where $m^0=5.50$ MeV and $\Lambda=0.631$ GeV. The dotted curve corresponds to the
 use of a constant value $G=5.691$\gev{-2}, in the notation of Ref. [12]. For the solid and dashed
 curves we have used $G(T)=G[1-0.17(T/T_c)]$. For the solid curve $T_c=0.170$ GeV and for
 the dashed curve we have used $T_c=0.150$ GeV.}
 \end{figure}

The organization of our work is as follows. In Sec.II we review
our calculations of polarization functions at finite temperature.
In Sec.III we discuss the calculation of hadronic current
correlation functions, making use of the results presented in
Sec.II. In Sections IV and V we present the results of our
numerical calculations of correlators of pseudoscalar and vector
currents, respectively. We also compare our results to some recent
lattice calculations of pseudoscalar and vector correlators
[14-16]. Finally, in Sec.VI we present same further discussion and
conclusions.

\begin{figure}
 \includegraphics[bb=0 0 280 235, angle=0, scale=1.2]{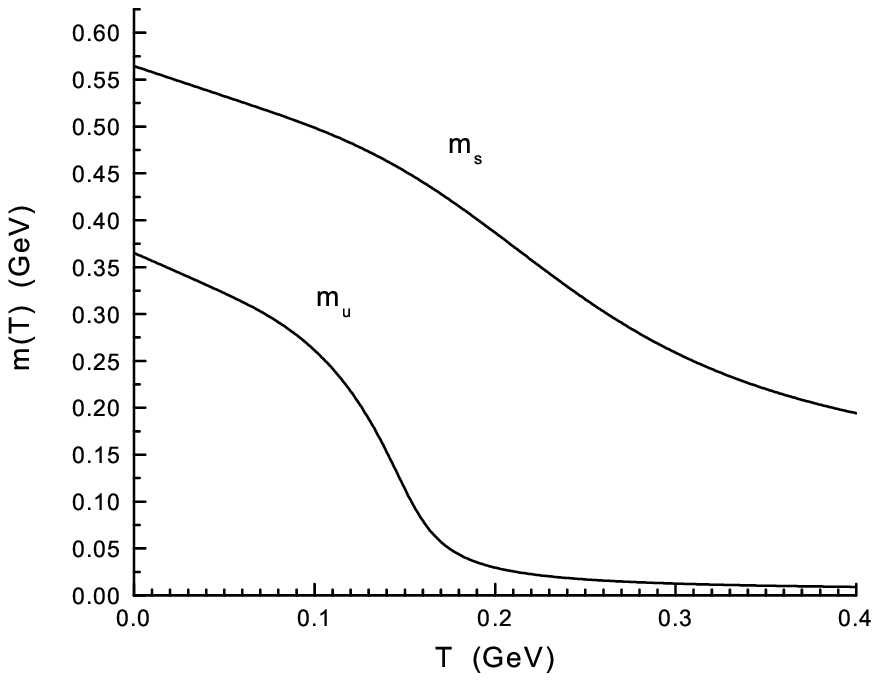}%
\caption{Temperature dependent constituent mass values, $m_u(T)$
and $m_s(T)$, calculated in a mean-field approximation [12], are
shown. [See Eq. (1.4)]. Here $m_u^0=0.0055$ GeV, $m_s^0=0.120$
GeV, and $G(T)=5.691[1-0.17(T/T_c)]$, if we use Klevansky's
notation [12]. (The value of $G_S$ of Eq. (1.1) is twice the value
of $G$ used in [12].)}
\end{figure}

\section{Polarization Functions at finite temperature}

The basic polarization function that is calculated in the NJL
model is shown in Fig.3. We will consider calculations of such
functions in the frame where $\vec P=0$. In our earlier work,
calculations were made after a confinement vertex was included.
That vertex is represented by the filled triangular region in
Fig.3. However, we here consider calculations for $T\geq 1.2\,T_c$
where confinement may be neglected. We will, however, use the
temperature-dependent mass values shown in Fig.2.

 \begin{figure}
 \includegraphics[bb=0 0 600 350, angle=0, scale=0.4]{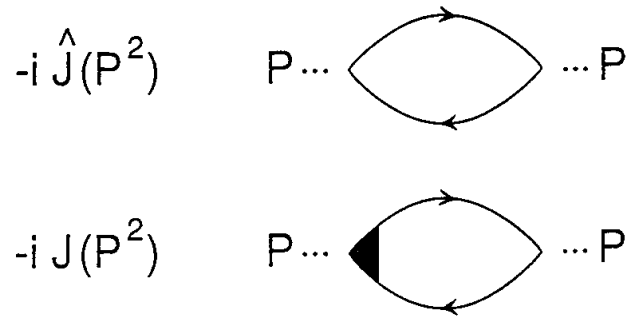}%
 \caption{The upper figure represents the basic polarization diagram of the
 NJL model in which the lines represent a constituent quark and a constituent
 antiquark. The lower figure shows a confinement vertex [filled triangular
 region] used in our earlier work. For the present work we neglect confinement
 for $T\geq1.2\,T_c$, with $T_c=150$ MeV.}
 \end{figure}

The procedure we adopt is based upon the real-time
finite-temperature formalism, in which the imaginary part of the
polarization function may be calculated. Then, the real part of
the function is obtained using a dispersion relation. The result
we need for this work has been already given in the work of Kobes
and Semenoff [17]. (In Ref. [17] the quark momentum in Fig.3 is
$k$ and the antiquark momentum is $k-P$. We will adopt that
notation in this section for ease of reference to the results
presented in Ref. [17].) With reference to Eq. (5.4) of Ref. [17],
we write the imaginary part of the scalar polarization function as
\be \mbox{Im}\,J_S(\textit{P}\,{}^2,
T)=\frac12(2N_c)\beta_S\,\epsilon(\textit{P}\,{}^0)\mytint
ke^{-\vec
k\,{}^2/\alpha^2}\left(\frac{2\pi}{2E_1(k)2E_2(k)}\right)\\\nonumber
\{(1-n_1(k)-n_2(k))
\delta(\textit{P}\,{}^0-E_1(k)-E_2(k))\\\nonumber-(n_1(k)-n_2(k))
\delta(\textit{P}\,{}^0+E_1(k)-E_2(k))\\\nonumber-(n_2(k)-n_1(k))
\delta(\textit{P}\,{}^0-E_1(k)+E_2(k))\\\nonumber-(1-n_1(k)-n_2(k))
\delta(\textit{P}\,{}^0+E_1(k)+E_2(k))\}\,.\ee Here, $E_1(k)=[\vec
k\,{}^2+m_1^2(T)]^{1/2}$. Relative to Eq. (5.4) of Ref.[17], we
have changed the sign, removed a factor of $g^2$ and have included
a statistical factor of $2N_c$, where the factor of 2 arises from
the flavor trace. In addition, we have included a Gaussian
regulator, $\exp[-\vec k\,{}^2/\alpha^2]$, with $\alpha=0.605$
GeV, which is the same as that used in most of our applications of
the NJL model in the calculation of meson properties. We also note
that \be n_1(k)=\frac1{e^{\,\beta E_1(k)}+1}\,,\ee and \be
n_2(k)=\frac1{e^{\,\beta E_2(k)}+1}\,.\ee For the calculation of
the imaginary part of the polarization function, we may put
$\ksq=m_1^2(T)$ and $(k-P)^2=m_2^2(T)$, since in that calculation
the quark and antiquark are on-mass-shell. In Eq. (2.1) the factor
$\beta_S$ arises from a trace involving Dirac matrices, such that
\be \beta_S&=&-\mbox{Tr}[(\slr k+m_1)(\slr k-\slr P+m_2)]\\
&=&2P^2-2(m_1+m_2)^2\,,\ee where $m_1$ and $m_2$ depend upon
temperature. In the frame where $\vec P=0$, and in the case
$m_1=m_2$, we have $\beta_S=2P_0^2(1-{4m^2}/{P_0^2})$. For the
scalar case, with $m_1=m_2$, we find \be \mbox{Im}\,J_S(P^2,
T)=\frac{N_cP_0^2}{4\pi}\left(1-\frac{4m^2}{P_0^2}\right)^{3/2}
e^{-\vec k\,{}^2/\alpha^2}[1-2n_1(k)]\,,\ee where \be \vec
k\,{}^2=\frac{P_0^2}4-m^2(T)\,.\ee

For pseudoscalar mesons, we replace $\beta_S$ by
\be \beta_P&=&-\mbox{Tr}[i\gamma_5(\slr k+m_1)i\gamma_5(\slr k-\slr P+m_2)]\\
&=&2P^2-2(m_1-m_2)^2\,,\ee which for $m_1=m_2$ is $\beta_P=2P_0^2$
in the frame where $\vec P=0$. We find, for the $\pi$ mesons, \be
\mbox{Im}\,J_P(P^2,T)=\frac{N_cP_0^2}{4\pi}\left(1-\frac{4m(T)^2}{P_0^2}\right)^{1/2}
e^{-\vec k\,{}^2/\alpha^2}[1-2n_1(k)]\,,\ee where $ \vec
k\,{}^2={P_0^2}/4-m_u^2(T)$, as above. Thus, we see that, relative
to the scalar case, the phase space factor has an exponent of 1/2
corresponding to a \textit{s}-wave amplitude. For the scalars, the
exponent of the phase-space factor is 3/2, as seen in Eq. (2.6).

For a study of the vector-isovector correlators, we introduce
conserved vector currents $j_{\mu,
i}(x)=\tilde{q}(x)\gamma_{\mu}\lambda_i q(x)$ with i=1, 2 and 3.
In this case we define \be J_V^{\mu\nu}(P^2,
T)=\left(g\,{}^{\mu\nu}-\frac{P\,{}^\mu
P\,{}^\nu}{P^2}\right)J_V(P^2, T)\ee and \be C_V^{\mu\nu}(P^2,
T)=\left(g\,{}^{\mu\nu}-\frac{P\,{}^\mu
P\,{}^\nu}{P^2}\right)C_V(P^2, T)\,,\ee taking into account the
fact that the current $j_{\mu,\,i}(x)$ is conserved. We may then
use the fact that \be J_V(P^2,T) =
\frac13g_{\mu\nu}J_V^{\mu\nu}(P^2,T)\ee and
\be\mbox{Im}\,J_V(P^2,T)&=&
\frac23\left[\frac{P_0^2+2m_u^2(T)}{4\pi}\right]
\left(1-\frac{4m_u^2(T)}{P_0^2}\right)^{1/2}e^{-\vec
k\,{}^2/\alpha^2}[1-2n_1(k)]\\ &\simeq&
\frac{2}{3}\mbox{Im}J_P(P^2,T)\,.\ee

We now consider \be \beta_{\mu\nu}^V=\mbox{Tr}[\gamma_\mu(\slr
k+m_1)\gamma_\nu(\slr k-\slr P+m_2)]\,,\ee and calculate \be
g^{\mu\nu}\beta_{\mu\nu}^V=4[P^2-m_1^2-m_2^2+4m_1m_2]\,,\ee which,
in the equal-mass case, is equal to $4P_0^2+8m^2(T)$, when
$m_1=m_2$ and $\vec P=0$. This result will be needed when we
calculate the correlator of vector currents in the next section.
Note that for the elevated temperatures considered in this work
$m_u(T)=m_d(T)$ is quite small, so that $4P_0^2+8m_u^2(T)$ can be
approximated by $4P_0^2$ when we consider the vector current
correlation functions. In that case, we have \be
\mbox{Im}\,J_V(P^2,T) \simeq
\frac{2}{3}\mbox{Im}\,J_P(P^2,T)\,,\ee At this point it is useful
to define functions that do not contain the Gaussian regulator:
\be\mbox{Im}\,\tilde{J}_P(P^2,T)=\frac{N_cP_0^2}{4\pi}\left(1-\frac{4m(T)^2}{P_0^2}\right)^{1/2}[1-2n_1(k)]\,,\ee
and
\be\mbox{Im}\,\tilde{J}_V(P^2,T)=\frac{2}{3}\frac{N_cP_0^2}{4\pi}\left(1-\frac{4m(T)^2}{P_0^2}\right)^{1/2}[1-2n_1(k)]\,.\ee
For the functions defined in Eq. (2.19) and (2.20) we need to use
a twice-subtracted dispersion relation to obtain
$\mbox{Re}\,\tilde{J}_P(P^2,T)$, or
$\mbox{Re}\,\tilde{J}_V(P^2,T)$. For example,
\be\mbox{Re}\,\tilde{J}_P(P^2,T)=\mbox{Re}\,\tilde{J}_P(0,T)+
\frac{P^2}{P_0^2}[\mbox{Re}\,\tilde{J}_P(P_0^2,T)-\mbox{Re}\,\tilde{J}_P(0,T)]+\\\nonumber
\frac{P^2(P^2-P_0^2)}{\pi}\int_{m^2(T)}^{\tilde{\Lambda}^{2}}
ds\frac{\mbox{Im}\,\tilde{J}_P(s,T)}{s(P^2-s)(P_0^2-s)}\,,\ee
where $\tilde{\Lambda}^{2}$ can be quite large since the integral
over the imaginary part of the polarization function is now
convergent. We may introduce $\tilde{J}_P(P^2,T)$ and
$\tilde{J}_V(P^2,T)$ as complex functions, since we now have both
the real and imaginary parts of these functions. We note that the
construction of either $\mbox{Re}\,J_P(P^2,T)$ or
$\mbox{Re}\,J_V(P^2,T)$ by means of a dispersion relation does not
require a subtraction. We use these functions to define the
complex functions $J_P(P^2,T)$ and $J_V(P^2,T)$.

In order to make use of Eq.(2.21) we need to specify
$\mbox{Re}\tilde{J}_P(0)$ and $\mbox{Re}\tilde{J}_P(P_0^2)$. We
found it useful to take $P_0^2=-1.0$ \gev2 and to put
$\mbox{Re}\tilde{J}_P(0)=\mbox{Re}J_P(0)$ and
$\mbox{Re}\tilde{J}_P(P_0^2)=\mbox{Re}J_P(P_0^2)$. The quantities
$\mbox{Re}\tilde{J}_V(0)$ and $\mbox{Re}\tilde{J}_V(P_0^2)$ are
determined in an analogous function. This procedure in which we
fix the behavior of a function such as $\tilde{J}_P(P^2)$ or
$\tilde{J}_V(P^2)$, which may be used when making calculations for
large $P^2$, is quite analogous to the procedure used in Ref.
[18]. In that work we made use of dispersion relations to
construct a continuous vector-isovector current correlation
function that had the correct perturbative behavior for
$P^2\rightarrow-\infty$ and also described the low-energy
resonance present in the correlator due to the excitation of the
$\rho$ meson. In Ref.\,[18] the NJL model was shown to provide a
quite satisfactory description of the low-energy resonant behavior
of the vector-isovector correlation function.

We may compare our expressions for $(1/\pi)\mbox{Im}J(P^2)$ with
those given in Eq.(2.7) of Ref.[19]. We find that the scalar and
pseudoscalar polarization functions are defined there with a sign
opposite to ours. However, since we have defined the pseudoscalar
current using the matrix $i\gamma_5$ rather than $\gamma_5$, which
was used in Ref.[19], our expression for the spectral function
agrees with that of Ref.[19] with respect to the sign in the
pseudoscalar case. In the case of the vector and axial-vector
spectral functions, we have the same sign convention as that of
Ref.[19]. We disagree with the expressions for the spectral
functions given for the scalar and axial-vector currents in
Ref.[19] with respect to the exponent of the phase-space factor.
In Ref.[19], the phase-space factor for the scalar and
axial-vector case is $[1-(2m/\omega)^2]^{1/2}$, while we have
$[1-(2m/\omega)^2]^{3/2}$ for those spectral functions. We agree
that the phase-space factor is $[1-(2m/\omega)^2]^{1/2}$ for the
pseudoscalar and vector current spectral functions.

In Ref.[19] use is made of the hard-thermal-loop (HTL)
approximation. It was found that the scalar channels are only
moderately influenced by the HTL medium effects, while the HTL
vertex corrections lead to divergent vector correlators.

\section{calculation of hadronic current correlation functions}
In this section we consider the calculation of
temperature-dependent hadronic current correlation functions. The
general form of the correlator is a transform of a time-ordered
product of currents, \be iC(P^2, T)=\int d^4xe^{iP\cdot
x}<\!\!<T(j(x)j(0))>\!\!>\,,\ee where the double bracket is a
reminder that we are considering the finite-temperature case.

For the study of pseudoscalar states, we may consider currents of
the form $j_{P,i}(x)=\tilde{q}(x)i\gamma_5\lambda^iq(x)$, where,
in the case of the $\pi$ mesons, $i=1,2$ and $3$. For the study of
scalar-isoscalar mesons, we introduce
$j_{S,i}(x)=\tilde{q}(x)\lambda^i q(x)$, with $i=0$ for the
flavor-singlet current and $i=8$ for the flavor-octet current [7].

In the case of the pseudoscalar-isovector mesons, the correlator
may be expressed in terms of the basic vacuum polarization
function of the NJL model, $J_P(P^2, T)$ [9,10,12]. Thus, \be
C_P(P^2, T)=J_P(P^2, T)\frac{1}{1-G_{P}(T)J_P(P^2, T)}\,,\ee where
$G_P(T)$ is the coupling constant appropriate for our study of
$\pi$ mesons. We have found $G_P(T)=13.49$\gev{-2} by fitting the
pion mass in a calculation made at $T=0$, with $m_u = m_d =0.364$
GeV. The result given in Eq. (3.2) is only expected to be useful
for small $P^2$, since the Gaussian regulator strongly modifies
the large $P^2$ behavior. Therefore, we suggest that the following
form is useful, if we are to consider the larger values of $P^2$
\be \frac{C_{P}(P^2, T)}{P^2}=\left[\frac{\tilde{J}_P(P^2,
T)}{P^2}\right] \frac{1}{1-G_P(T)J_P(P^2, T)}\,.\ee (As usual, we
put $\vec{P}=0$.) This form has two important features. At large
$P_0^2$, ${\mbox{Im}\,C_{P}(P_0, T)}/{P_0^2}$ is a constant, since
${\mbox{Im}\,\tilde{J}_{P}(P_0^2, T)}$ is proportional to $P_0^2$.
Further, the denominator of the second term on the right-hand side
of Eq.(3.3) goes to 1 for large $P_0^2$. On the other hand, at
small $P_0^2$, that denominator is capable of describing resonant
enhancement of the correlation function. (\,We may again refer to
Ref. [18], in which a similar approximation is described.)

We then have \be
C_V(P^2,T)=\tilde{J}_V(P^2,T)\frac1{1-G_V(T)J_V(P^2,T)}\,,\ee
where we have introduced \be\mbox{Im}\tilde{J}_V(P^2,T)&=&
\frac23\left[\frac{P_0^2+2m_u^2(T)}{4\pi}\right]
\left(1-\frac{4m_u^2(T)}{P_0^2}\right)^{1/2}[1-2n_1(k)]\\
&\simeq& \frac{2}{3}\mbox{Im}\tilde{J}_P(P^2,T)\,. \ee (See
Eq.(2.7) for the specification of $k=|\vec k|$.) In the
literature, $\omega$ is used instead of $P_0$ [\,14 -16\,]. We may
define the spectral functions \be\sigma_V(\omega,
T)=\frac{1}{\pi}\,\mbox{Im}\,C_V(\omega, T)\,,\ee and
\be\sigma_P(\omega, T)=\frac{1}{\pi}\,\mbox{Im}\,C_P(\omega,
T)\,.\ee

We may use the notation $\bar \sigma_P(\omega, T)$ and $\bar
\sigma_V(\omega, T)$ for the spectral functions given in the
literature [14-16]. We have the following relations: \be \bar
\sigma_P(\omega, T)=\sigma_P(\omega, T)\ee and \be
\frac{\bar\sigma_V(\omega, T)}2=\frac 3 4\sigma_V(\omega, T)\,,\ee
where the factor of 3/4 arises because, in Refs. [14-16], there is
a division by 4, while we have divided by 3 as in Eq. (2.13).

In addition to providing values of $\sigma_V(\omega, T)/\omega^2$
and $\sigma_P(\omega, T)/\omega^2$, the $\tau$-dependent hadronic
correlators $G_P(\tau, T)$ and $G_V(\tau, T)$ were introduced
[\,14-16\,], with \be G(\tau, T)=\int_0^\infty d\omega
\sigma(\omega, T)K(\omega, \tau, T)\,.\ee Here, \be K(\omega,
\tau, T) =
\frac{\cosh[\,\omega(\tau-\frac{1}{2T})\,]}{\sinh(\frac{\omega}{2T})}\,.\ee

We will present results of our calculations of $G_P(\tau, T)$ and
$G_V(\tau, T)$ in Sections IV and V, respectively. We will also
present the values obtained for $\mbox{Im}\,C_P(P_0, T)/P_0^2$ and
$\mbox{Im}\,C_V(P_0, T)/P_0^2$ for various values of $T/T_c$ in
Sections IV and V.

We note that values have been obtained for the spectral functions
in the scalar, vector, pseudoscalar and axial-vector channels in
Ref.[20]. These functions exhibit resonant features at about 2, 7
and 14 GeV when $T\approx 1.4T_c$. High-energy resonances are
still present at 5 and 13 GeV, when $T\approx 1.9T_c$. The
interpretation of these high-energy resonances is uncertain [21].

\section{correlation functions for pseudoscalar-isovector currents - numerical results}
In Fig.4 we show $\mbox{Im}\,C_P(P_0, T)/P_0^2$ for $T/T_c=1.5$,
and for model 1, as a solid line. The dashed line represents the
result at $T/T_c=1.5$, if we put $G_P(T)=0$. The results for a
broader range of temperatures are shown in Fig.5. In Fig.6 we show
the ratios $R_P(P_0, T)=\mbox{Im} C_P(P_0, T)/\mbox{Im}
C_P^{(0)}(P_0, T)$ calculated for various temperatures. Here,
$\mbox{Im} C_P^{(0)}(P_0, T)/P_0^2$ is calculated with $G_P(T)=0$.
In Fig.7 we show the ratio $R_G^{(P)}(\tau, T)=G_P(\tau,
T)/G_P^{(0)}(\tau, T)$ for various temperatures. Here,
$G_P^{(0)}(\tau, T)$ represents $G_P(\tau, T)$ evaluated for
$G_P(T)=0$. In Fig.8 we exhibit $\mbox{Im}\,C_P(P_0, T)/P_0^2$ for
various temperature for model 2.

 \begin{figure}
 \includegraphics[bb=0 0 280 220, angle=0, scale=1]{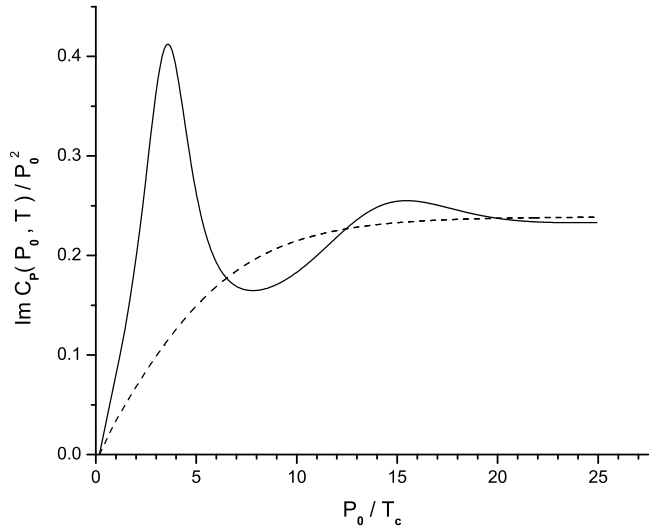}%
 \caption{Values of $\mbox{Im}\,C_P(P_0, T)/P_0^2$ are shown at $T/T_c=1.5$
 for model 1, where $G_P(T)=G_P[1-0.17\,(T/T_c)]$ (solid line). The dashed line
 represents the result obtained when $T/T_c=1.5$ and $G_P(T)=0$. (Here, the dashed line
 represents the values of $\mbox{Im}\,\tilde{J}_P(P_0, T)/P_0^2$ for $T/T_c=1.5$.)}
 \end{figure}

 \begin{figure}
 \includegraphics[bb=0 0 280 220, angle=0, scale=1]{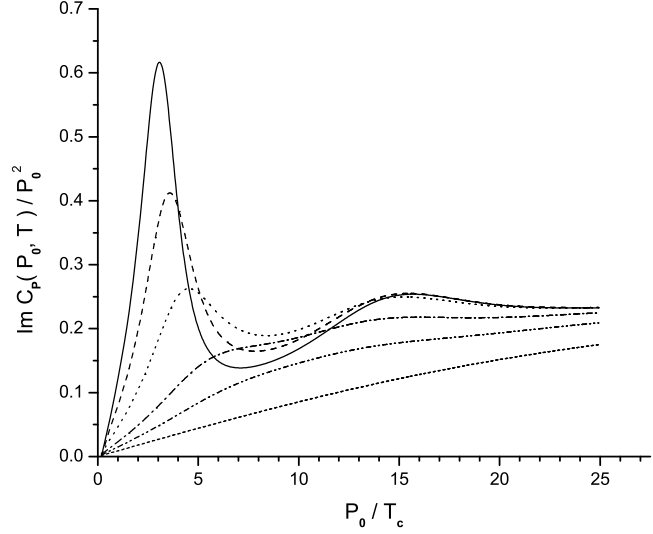}%
 \caption{Values of $\mbox{Im}\,C_P(P_0, T)/P_0^2$ are shown for model 1 and for various
 temperatures: $T/T_c=1.2$ [sold line], $T/T_c=1.5$ [dashed line], $T/T_c=2.0$ [dotted line],
 $T/T_c=3.0$ [dashed-dotted line], $T/T_c=4.0$ [dashed-double dotted line], and $T/T_c=5.88$ [short-dashed line].}
 \end{figure}

\begin{figure}
 \includegraphics[bb=0 0 280 220, angle=0, scale=1]{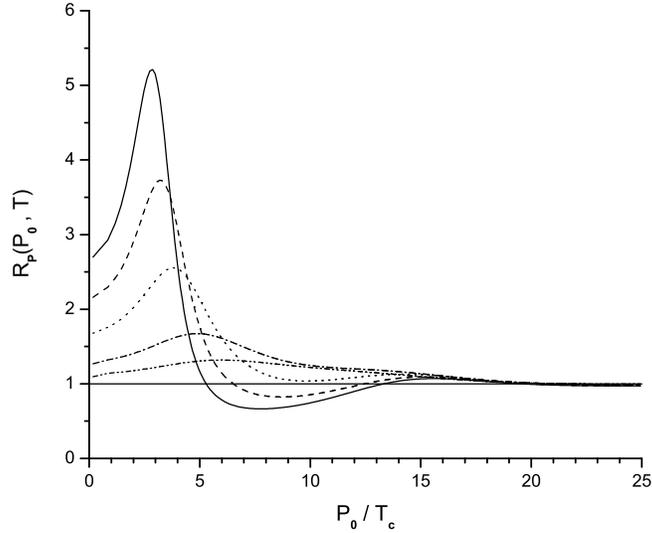}%
 \caption{The ratio $R_P(P_0, T)=\mbox{Im} C_P(P_0, T)/\mbox{Im}
C_P^{(0)}(P_0, T)$ is shown for various values of $T/T_c=1.2$
[solid line], $T/T_c=1.5$ [dashed line], $T/T_c=2.9$ [dotted
line], $T/T_c=3.0$ [dot-dashed line] and $T/T_c=4.0$ [double
dot-dashed line]. Here, $\mbox{Im} C_P^{(0)}(P_0, T)$ is
calculated at temperature $T$ with $G_P(T)=0$.}
 \end{figure}

\begin{figure}
 \includegraphics[bb=0 0 280 220, angle=0, scale=1]{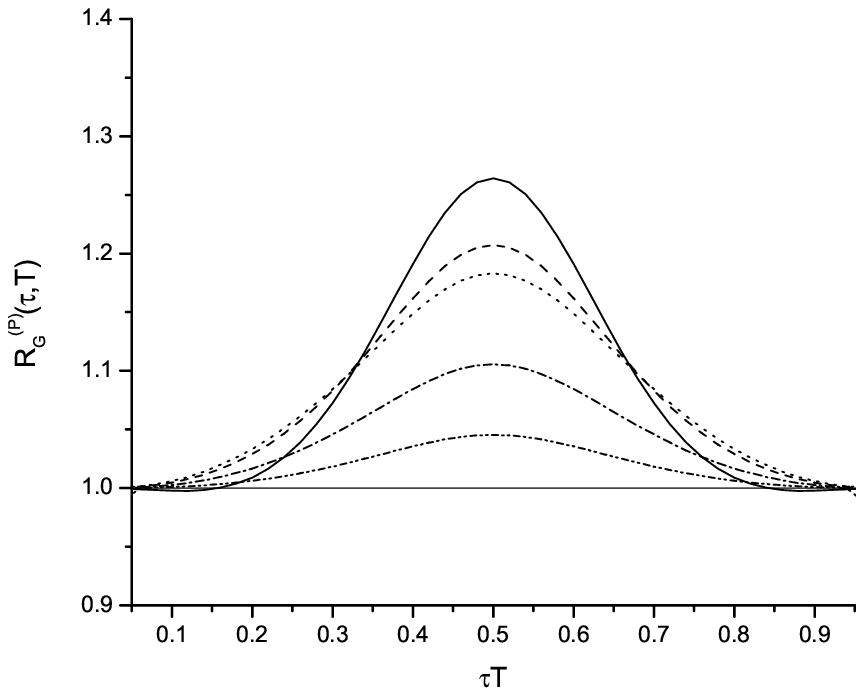}%
 \caption{Values of $R_G^{(P)}(\tau,T)$ are shown for values of $T/T_c=1.2$ [solid line], $T/T_c=1.5$ [dashed line],
 $T/T_c=2.0$ [dotted line],
 $T/T_c=3.0$ [dashed-dotted line], and $T/T_c=4.0$ [dashed-double dotted line]. For $T/T_c=5.88$, $R_G^{(P)}
 (\tau, T)=1$.}
\end{figure}

In Fig.9 we show, as a dashed line, values of $\mbox{Im}\,C_P(P_0,
T)/P_0^2$ at $T/T_c=5.88$ obtained using a constant value of
$G_P(T)=13.49$ \gev{-2}. The solid line is the result for model 1
(or model 2) at $T/T_c=5.88$, in which case $G_P(T)=0$ for both
models. Since we have argued that for $T/T_c\sim6$ the quark-gluon
plasma should be a weakly interacting system, the use of a
constant value for $G_P(T)$ appears to be unacceptable. We present
this result to support our argument that the coupling parameters
of the chiral Lagrangian model should be made temperature
dependent to be consistent with QCD thermodynamics.

\begin{figure}
 \includegraphics[bb=0 0 280 220, angle=0, scale=1]{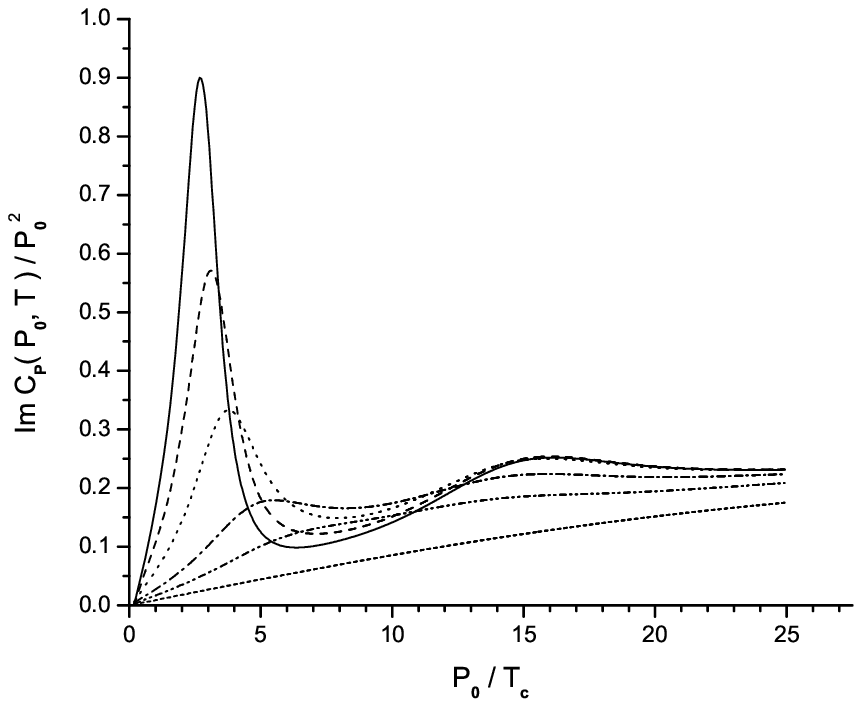}%
 \caption{
 Values of $\mbox{Im}\,C_P(P_0, T)/P_0^2$ are shown at for various
 temperatures for model 2, with $G_P(T)=G_P[1-0.0289\,(T/T_c)^2]$. (See the caption to
 Fig.5.)}
 \end{figure}

 \begin{figure}
 \includegraphics[bb=0 0 280 220, angle=0, scale=1]{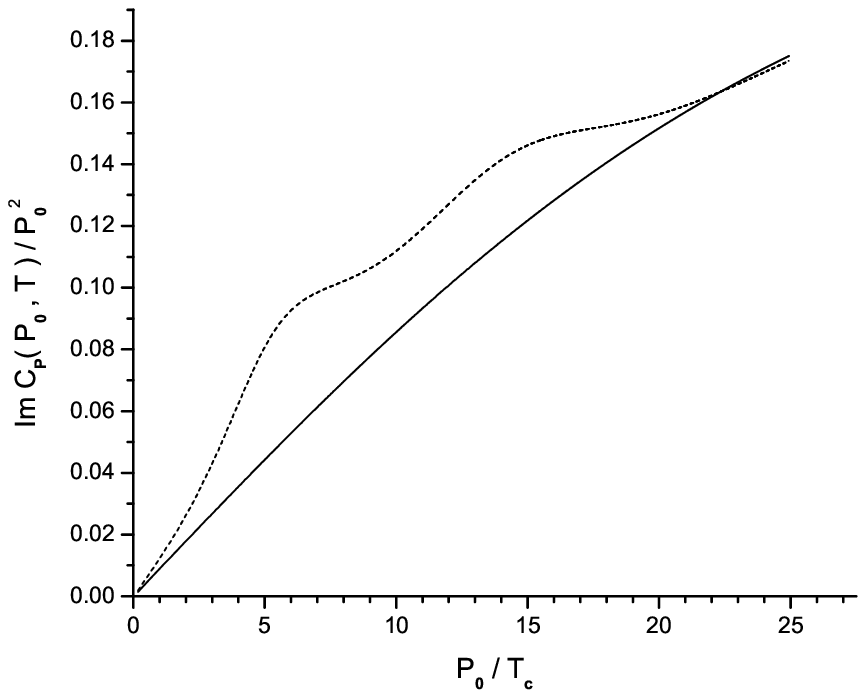}%
 \caption{Values of $\mbox{Im}\,C_P(P_0, T)/P_0^2$ are shown for $T/T_c=5.88$.
 The dashed line represents the result when $G_P(T)=G_P=13.49$\gev{-2}. The solid
 line is the result for $G_P(T)=0$, which is characteristic of models 1 and 2,
 when $T/T_c=5.88$. (The solid line, therefore, represents the values of
 $\mbox{Im}\,\tilde{J}_P(P_0, T)/P_0^2$ for $T/T_c=5.88$.)}
 \end{figure}

In Fig.10 we show values of $G_P(\tau, T)/T^3$ for $T/T_c=1.5$
[dashed line], $T/T_c=3.0$ [dash-dotted line] and $T/T_c=5.88$
[solid line]. In Fig.1 of Ref. [11] we see a marked difference in
the behavior of $G_P(\tau, T)/T^3$ and $G_V(\tau, T)/T^3$. (We
will present some discussion of this matter in Sec.V.)

 \begin{figure}
 \includegraphics[bb=0 0 280 220, angle=0, scale=1]{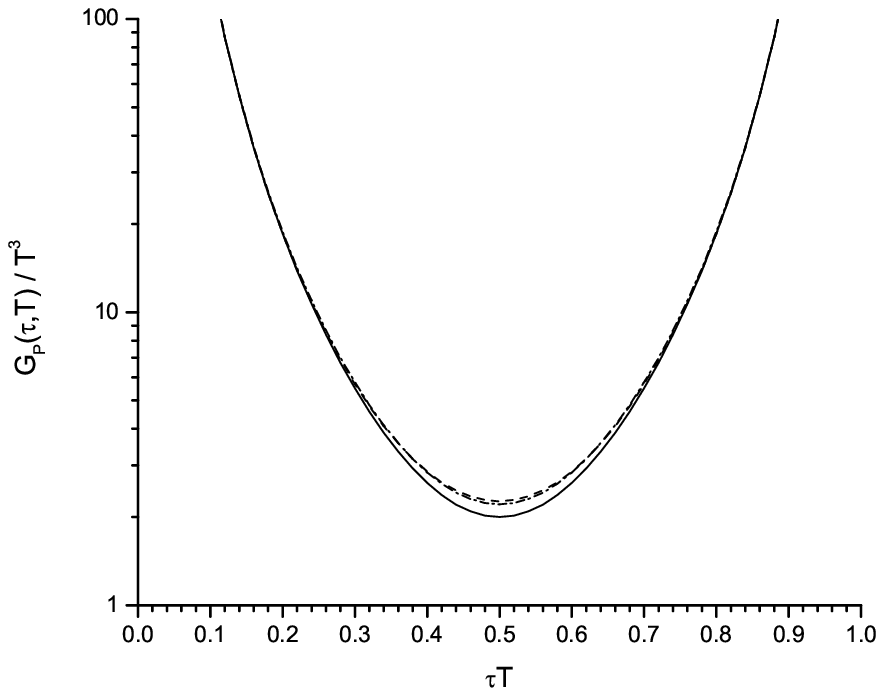}%
 \caption{Values of $G_P(\tau, T)/T^3$ are shown for $T/T_c=1.5$ [dashed line],
 $T/T_c=3.0$ [dashed-dotted line], and $T/T_c=5.88$ [solid line]. These results
 were calculated with model 1.}
 \end{figure}

\section{correlation functions for vector-isovector currents - numerical results}

In Fig.11 we show the values of $\mbox{Im}\,C_V(P_0, T)/P_0^2$ for
model 1 and for $T/T_c=1.5$, as a solid line. The result for
$T/T_c=1.5$ and $G_V(T)=0$ is represented by the dashed line. [See
Fig.4.] In Fig.12 we show values of $\mbox{Im}\,C_V(P_0, T)/P_0^2$
for various values of $T/T_c$ and for model 1. Corresponding
results for model 2 are given in Fig.13.

 \begin{figure}
 \includegraphics[bb=0 0 280 220, angle=0, scale=1]{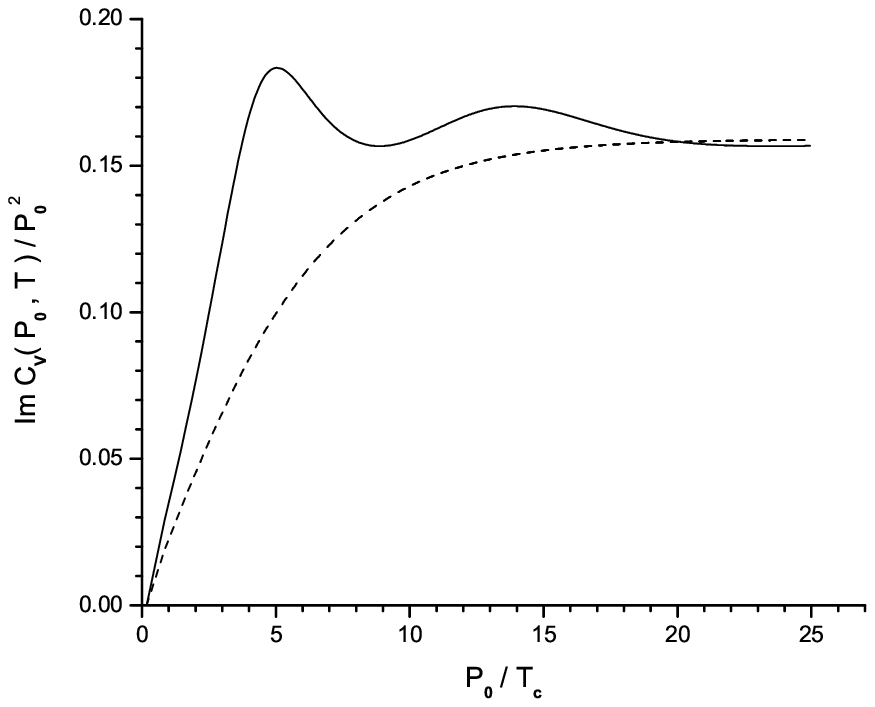}%
 \caption{Values of $\mbox{Im}\,C_V(P_0, T)/P_0^2$ are shown for $T/T_c=1.5$.
 (See the caption of Fig.4.)}
 \end{figure}

 \begin{figure}
 \includegraphics[bb=0 0 280 220, angle=0, scale=1]{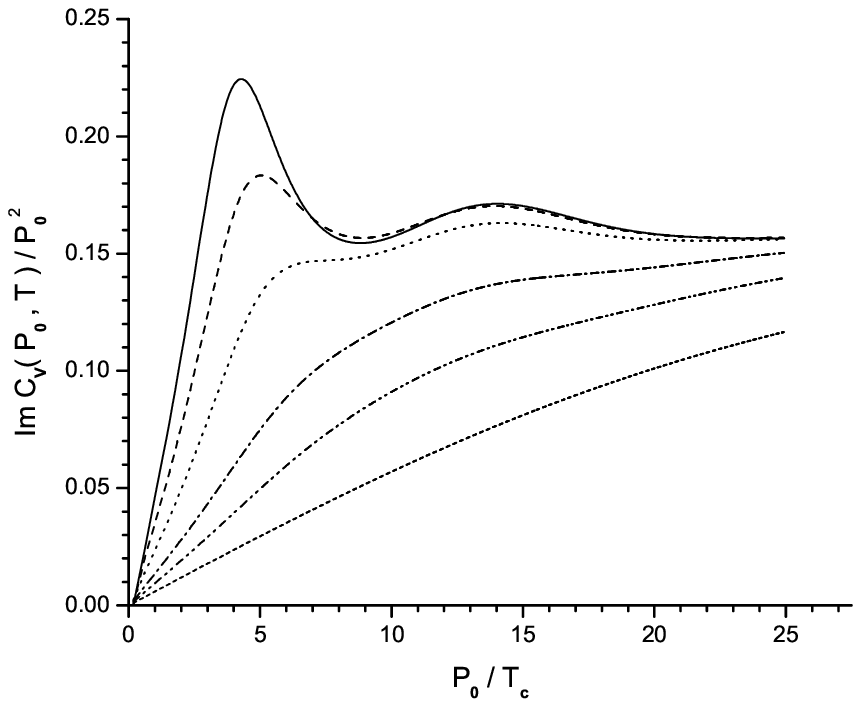}%
 \caption{Values of $\mbox{Im}\,C_V(P_0, T)/P_0^2$ are shown for model 1
 and for various temperatures. (See the caption of Fig.5.)}
 \end{figure}

In Fig.14 we show the ratio $R_V(P_0, T)=\mbox{Im} C_V(P_0,
T)/\mbox{Im} C_V^{(0)}(P_0, T)$, where $ \mbox{Im} C_V^{(0)}(P_0,
T)$ is calculated with $G_V(T)=0$. (See Fig.9). In Fig.15 we show
the ratio $R_G^{(V)}(\tau, T)=G_V(\tau, T)/G_V^{(0)}(\tau, T)$,
where $G_V{(0)}(\tau, T)$ represents $G_V(\tau, T)$ calculated
with $G_V(T)=0$. In Fig.16 we show, as a dashed line, values for
$T/T_c=5.88$ of $\mbox{Im}\,C_V(P_0, T)/P_0^2$, which were
calculated with a constant value of $G_V(T)=11.46$ \gev{-2}. The
solid line represents the results for $G_V(T)=0$, which is
characteristic of models 1 and 2 when $T/T_c=5.88$. The comments
made with respect to Fig.9 are also applicable here.

 \begin{figure}
 \includegraphics[bb=0 0 280 220, angle=0, scale=1]{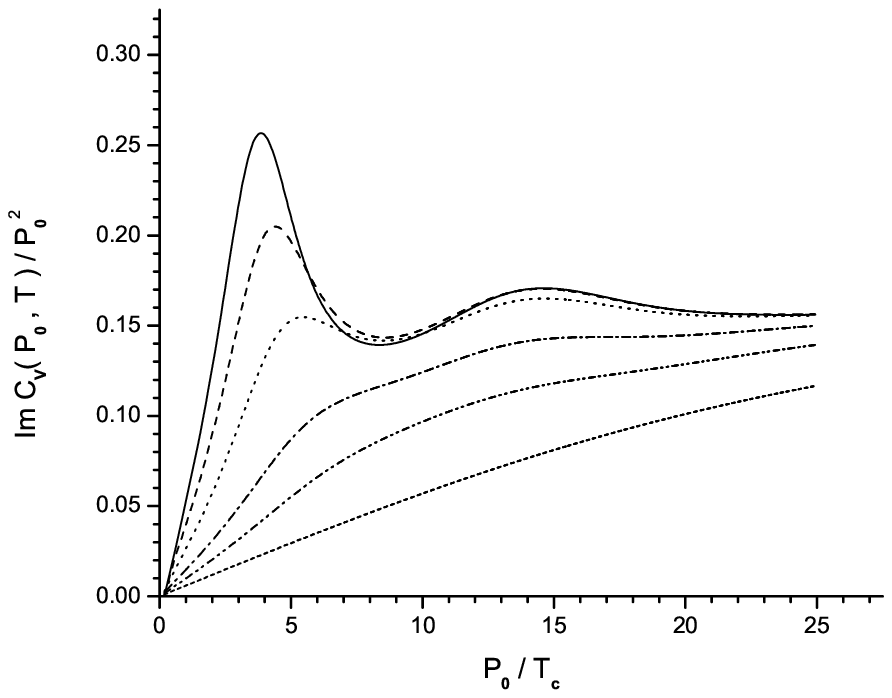}%
 \caption{Values of $\mbox{Im}\,C_V(P_0, T)/P_0^2$ are shown for model 2
 and for various temperatures. (See the caption of Fig.5.) }
 \end{figure}

 \begin{figure}
 \includegraphics[bb=0 0 280 220, angle=0, scale=1]{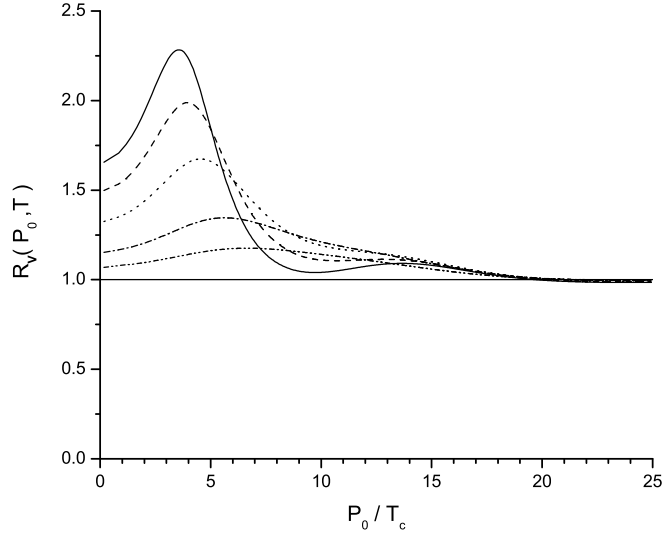}%
 \caption{Same caption as Fig.6, except that we show $R_V(P_0, T)=\mbox{Im} C_V(P_0, T)/\mbox{Im}
C_V^{(0)}(P_0, T)$. }
 \end{figure}

\begin{figure}
 \includegraphics[bb=0 0 280 220, angle=0, scale=1]{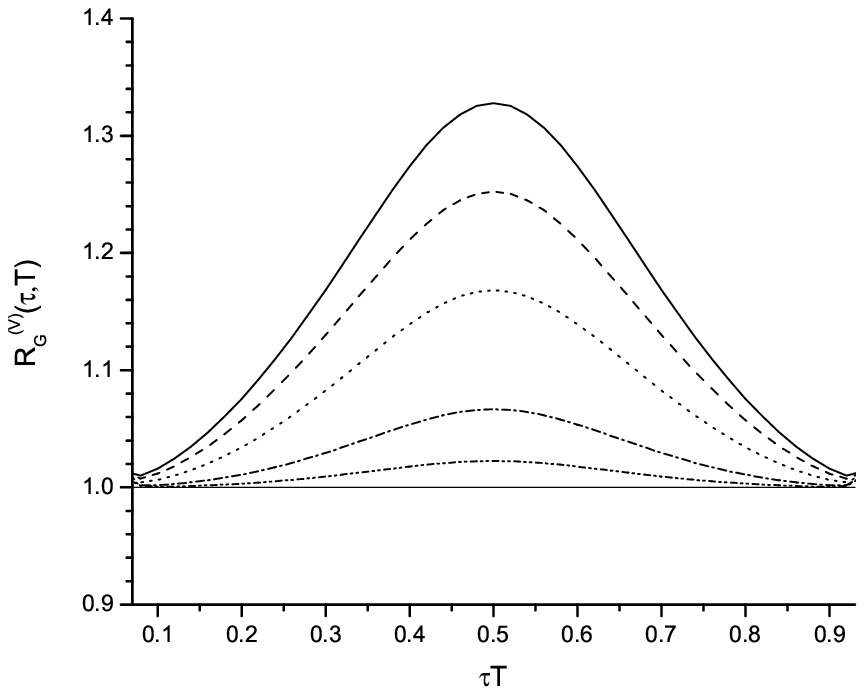}%
 \caption{Values of $R_G^{(V)}(\tau,T)$ are shown for values of $T/T_c=1.2$ [solid line], $T/T_c=1.5$ [dashed line],
 $T/T_c=2.0$ [dotted line],
 $T/T_c=3.0$ [dashed-dotted line], and $T/T_c=4.0$ [double dot-dashed line]. For $T/T_c=5.88$, $R_G^{(V)}(\tau, T)=1$.}
 \end{figure}

\begin{figure}\includegraphics[bb=0 0 280 220, angle=0, scale=1]{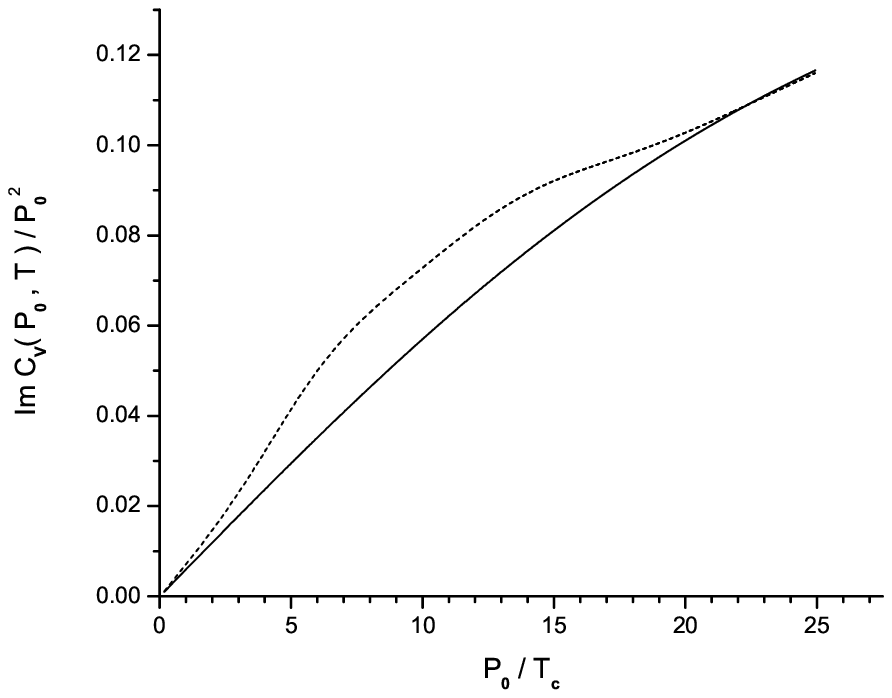}%
 \caption{Values of $\mbox{Im}\,C_V(P_0, T)/P_0^2$ are shown for $T/T_c=5.88$.
 The dashed line represents the result when $G_V(T)=G_V=11.46$\gev{-2}. The solid
 line is the result for $G_V(T)=0$, which is characteristic of models 1 and 2,
 when $T/T_c=5.88$. The solid line, therefore, represents the values of
 $\mbox{Im}\,\tilde{J}_V(P_0, T)/P_0^2$ for $T/T_c=5.88$.}
 \end{figure}

In Fig.17 we show values of $G_V(\tau, T)/T^3$ for $T/T_c=1.5$
[dashed line], $T/T_c=3.0$ [dash-dotted line] and $T/T_c=5.88$
[solid line]. These results were obtained with model 1, and may be
compared to those shown in Fig.9, where we see generally similar
behavior. That is in strong contrast to the results shown in Ref.
[15], where $G_V(\tau, T)/T^3$ and $G_P(\tau, T)/T^3$ shows quite
different behavior, with the result for the vector correlator
close to the values for $G(T)=0$ at $T/T_c=1.5$ and $T/T_c=3.0$.
This suggests that the value of $G_V=11.46$\gev{-2} that we have
used in this work may be too large, or that the temperature
dependence of $G_V(T)$ is such as to yield smaller values than
those obtained in this work for model 1 or model 2. (In general,
we prefer the the second of these possibilities, since our value
of $G_V$ at $T=0$ is set by fitting meson spectra.) In Fig.18 we
show $\mbox{Im}\,C_V(P_0, T)/P_0^2$ for $T=1.5\,T_c$, for model 1,
with $G_V=11.46$\gev{-2} [solid line], $G_V=8.00$\gev{-2} [dashed
line], $G_V=4.0$\gev{-2} [dotted line] and $G_V=0.0$\gev{-2}
[dot-dashed line]. It would be of interest to consider the results
of Ref. [15] for the Euclidean-space correlator to be correct and
to then determine $G_V(T)$ so that we fit the values of $G_V(\tau,
T)/T^3$ obtained in the lattice calculations. We defer such a
project to a future work.

 \begin{figure}
 \includegraphics[bb=0 0 280 220, angle=0, scale=1]{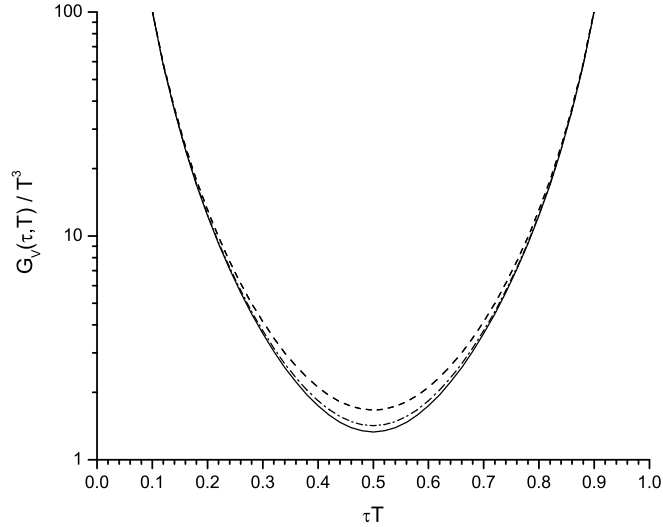}%
 \caption{Values of $G_V(\tau, T)/T^3$ are shown for
 $T/T_c=1.5$ [dashed line],  $T/T_c=3.0$ [dashed-dotted line], and
 $T/T_c=5.88$ [solid line]. These results were calculated with model 1.}
 \end{figure}

 \begin{figure}
 \includegraphics[bb=0 0 280 220, angle=0, scale=1]{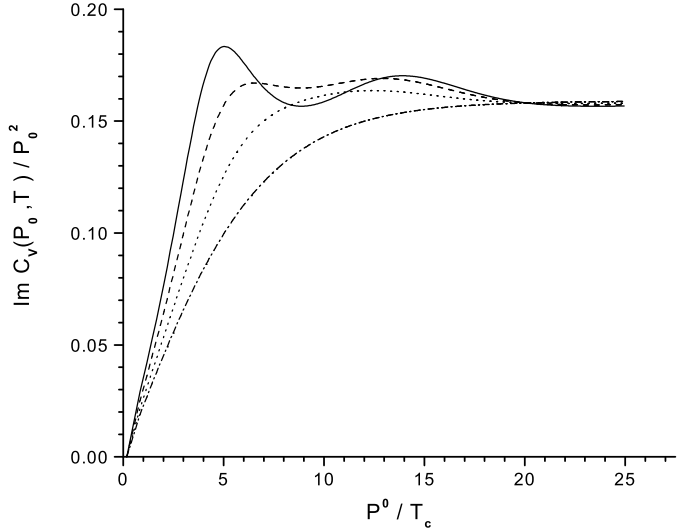}%
 \caption{Values of $\mbox{Im}\,C_V(P_0, T)/P_0^2$ are shown for
 $T/T_c=1.5$ with $G_V=11.46$\gev{-2} [solid line], $G_V=8.00$\gev{-2} [dashed line],
 $G_V=4.00$\gev{-2} [dotted line] and $G_V=0.0$\gev{-2} [dot-dashed line].
 Here, we have used model 1.}
 \end{figure}

\section{discussion}

It is difficult to make a definitive comparison of our results and
the results obtained for the spectral functions in the lattice
simulations, since those results are accompanied by large errors.
Also, it is somewhat difficult to understand why the different
behavior seen for $G_V(\tau, T)/T^3$ and $G_P(\tau, T)/T^3$ in
Ref. [15], leads to values for $\sigma_P(\omega, T)/\omega^3$ and
$\sigma_V(\omega, T)/\omega^3$ that are rather similar [14-16]. On
the other hand, the chiral Lagrangian model provides a systematic
study, which may yield some guidance for further studies of
lattice QCD. This may be particularly important in the light of
the comments made in Ref. [15]:

\hspace{1cm} \begin{quote} "The reconstruction of
$\sigma_H(\omega, T)$ and in particular the determination of its
low energy structure thus is difficult at non-zero temperature.
Additional complications arise in lattice calculations which
necessarily are performed on lattices with finite number of points
$(N_{\tau})$ in Euclidean time. The correlation functions
$G_H(\tau, T)$ can thus be calculated only at finite set of
Euclidean times $\tau T=k/N_{\tau}$, with $k=0, ...N_{\tau}-1$. In
order to reconstruct the spectral functions from this limited set
of information it is necessary to include in the statistical
analysis of numerical results also prior information on the
structure of $G_H(\tau, T)$ as well as assumptions about the
likelihood of a certain spectral function $\sigma_H(\omega, T)$.
It has been suggested to provide this additional information
through the application of the Maximum Entropy Method (MEM) [22,
23], which has been applied successfully to many other
ill-conditioned problems in physics $\cdots$."
\end{quote}

While it may be premature to compare our results for the spectral
functions with those given in literature, the errors for the
Euclidean-time correlation functions are small. Therefore, in Ref.
[24] we have made an extensive comparison of our results with
those presented in Ref. [15]. In that work we provided additional
evidence for the temperature dependence of the coupling parameters
that we have introduced.




\end{document}